# Reverse energy flows: the physical mechanism underling dramatic reduction of loss in hollow-core fibers

Andrey Pryamikov[1] and Sergei Turitsyn[2]

[1]Prokhorov General Physics Institute of Russian Academy of Sciences, Moscow, Russia

Email Address: pryamikov@mail.ru

[2]Aston Institute of Photonic Technologies Aston University, Birmingham, UK

Email Address: s.k.turitsyn@aston.ac.uk

**Abstract.** Hollow-core fibers (HCFs) with claddings composed of silica glass capillaries have recently attracted significant attention following the demonstration of optical losses lower than in conventional telecommunication fibers. It is well established that optical losses in HCFs are highly sensitive to both the wavelength and the geometry of the cladding capillaries. The underlying physical mechanisms behind reducing loss with the change of HCF design parameters while keeping the same fiber structure are still not fully understood. We show here that the shape of the capillaries plays a crucial role in controlling radial energy backflows that influence light confinement and the energy leakage from air-core modes of HCFs. Through numerical modeling, we demonstrate that optimizing the capillary geometry to tailor the distribution of reverse radial energy fluxes leads to a substantial reduction in transmission losses. Observed occurrences of vortex of the Poynting vector allows us to draw an interesting interdisciplinary analogy with the hydrodynamical system with suppressed backward flow - Tesla valve. The combination of singular optics and energy flux analysis provides valuable physical insight into the mechanisms governing waveguiding in HCFs offering a pathway toward novel designs with minimized leakage loss.

# 1. Introduction

Single-mode and multi-mode optical fibers, including silica-based, polymer, micro-structured and others, are widely used in a range of applications from telecommunications and sensing to imaging and lasers. Though the idea of guiding light in hollow tubes dates back to 19th century [1], the renewed surge of interest in hollow-core fibers is due to matured technology that not only offers low latency, but also low losses and low nonlinearity (see e.g. [2, 3, 4, 5, 6, 7, 8, 9, 10] and numerous references therein). General mechanisms of waveguiding and corresponding light localization in solid-core and hollow-core micro-structured fibers are well understood. In the case of solid-core fibers it is the principle of total internal reflection and its modification with the formation of photonic band gaps [11, 12, 13, 14]. In the case of hollow-core fibers, in addition to the formation of photonic band gaps [15], an antiresonance mechanism [16, 17] and inhibited coupling mechanism [18] are used to explain the mechanism of light localization in the air-core. In the simplified ARROW model, the cladding capillary wall is considered a planar Fabry- Perot resonator with corresponding maxima and minima of the transmission for radiation f lowing from the air-core that provides useful physical insight into the wave-guiding mechanism. Fiber losses can be analyzed within the inhibited coupling model, that considers an overlap integrals between the air core and the cladding modes. The minimum of such integrals corresponds to the minimum of losses for the hollow-core fiber. It is also known that strong light localization in the air-core can be achieved without two-dimensional photonic crystal structure in the fiber cladding. One layer of capillaries in the cladding can enable localization of radiation [7, 19] even in the mid-IR spectral range, where silica glass has large material losses. A curved shape of the core-cladding boundary is of critical importance, otherwise the effect of light localization is greatly reduced [7, 20, 19, 21]. The effective suppression of energy leakage from air-core modes occurs not only at the curved interfaces of the cladding capillaries, but also within the areas between them. Notably, it suffices to add additional nested capillaries in the cladding [22, 2] to substantially reduce losses down to the values comparable with the losses of telecommunication fibers. In [23, 24] the impact of the shape of the cladding capillary on the loss level of a hollow-core fiber was studied. A hollow core fiber using anisotropic anti-resonant tubes elongated in the radial direction in the cladding was proposed in [23] to decrease the loss while providing effectively single-mode guidance. Despite understanding the importance of the anti-resonance structure for light localization in HCFs, it is not entirely clear why in such anti-resonance fibers for some set of parameters, losses are dramatically reduced. The main goal of this work is to provide new insights on this dramatic reduction of loss at certain parameters within the visibly similar design of the anti-resonance fibers. We examine the connection between the shape of the core-cladding boundary of a HCF and the behavior of the radial component of the Poynting vector of the air-core modes. We demonstrate that by the design of the shape of the cladding capillaries one can change the distributions of reverse radial energy flows that will reduce energy leakage from the air-core modes of HCFs. Therefore, control of reverse radial energy flows can lead to a dramatic reduction of loss in the hollow–core fibers.

The reverse radial and azimuthal energy flows have been studied and demonstrated recently in the context of OAM beams [25, 26]. Consideration of reverse radial energy fluxes has been successfully applied to explain the so-called iso-propagation vortex beams with OAM-independent propagation dynamics. This approach employs the energy redistribution mechanism to reverse the radial energy flows of traditional vortex beams. Such vortex beams maintain resilience in diverse environments; for example, they experience reduced modal scattering in atmospheric turbulence. Consideration of the energy flows in HCF makes it possible to find some similarities with the hydrodynamics. We will draw here an interesting analogy between the light energy behavior in hollow-core fibers and fluid propagation in the hydrodynamic passive device, known as Tesla valve.

The paper is organized as follows. In Section 2, Subsection 2.1, we consider the change in the structure of the radial energy flux of the air-core mode and its leakage losses depending on the shape of the cladding capillaries or in the presence of nested cladding capillaries. In Subsection 2.2, we examine the behavior of the zero-level lines for the electric field components of the fundamental air-core mode and their relationship with the formation of phase edge dislocations in the walls of the cladding capillaries. A similar consideration is made for the zero-level lines of the projections of the transverse component of the Poynting vector of the air-core mode, and the dependence of losses on their shape is demonstrated. Section 3 contains a Discussion and Conclusions.

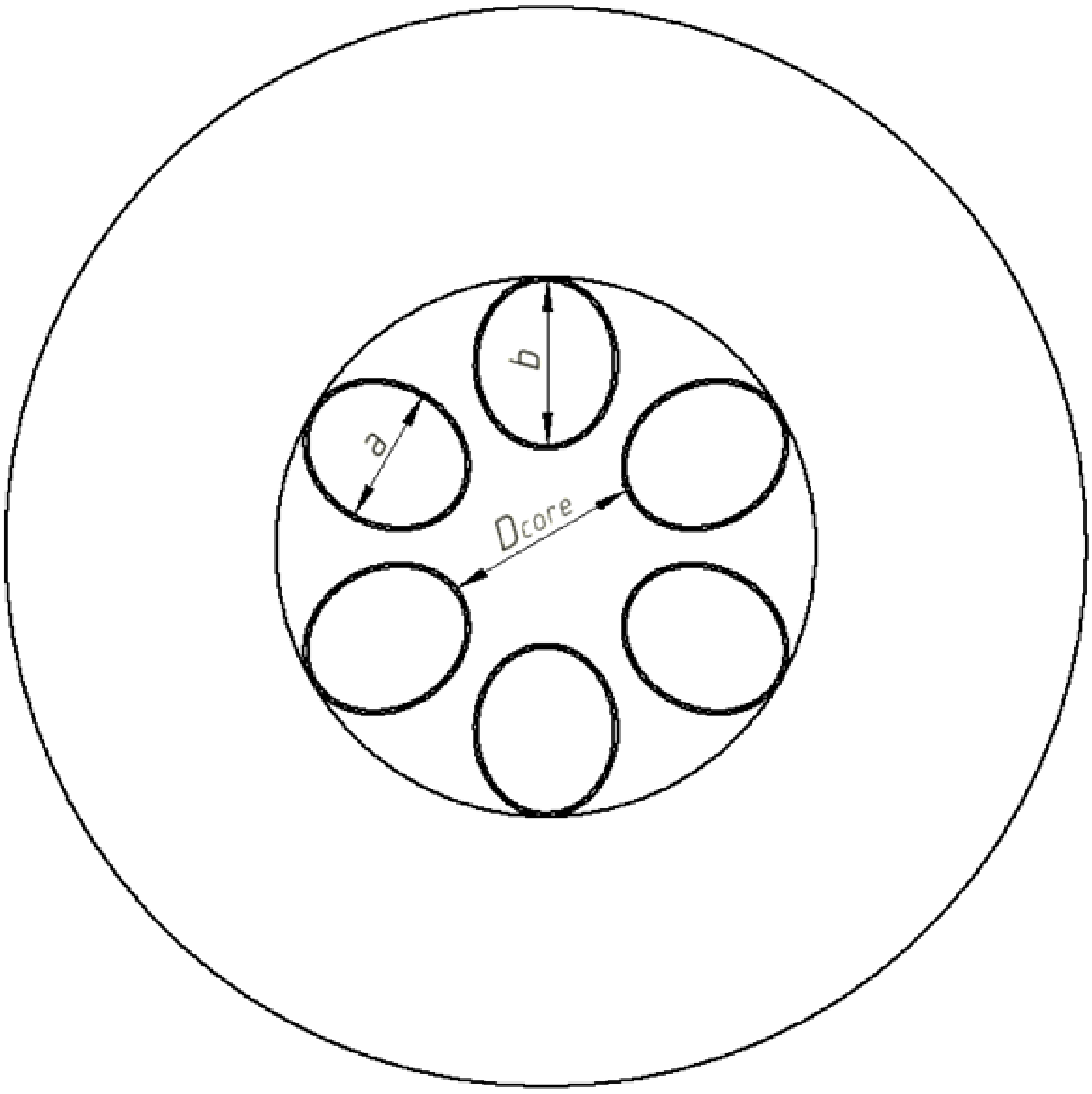

Figure 1: Cross-section of a hollow- core fiber with elliptical capillaries in the cladding

# 2. Results

## 2.1 The leakage loss dependence on the shape of the cladding capillary and the structure of energy flows in the capillary wall

In order to demonstrate the influence of the shape of the cladding capillary on the distribution of the radial energy flow $P_r(r,\phi)$ of the air - core mode in the capillary wall, we consider a hollow- core fiber with six elliptical capillaries in the cladding (Fig. 1). The calculation was carried out using the finite element method in the Comsol Multiphysics package. In order not to change the air- core diameter of the hollow- core fiber and "the anti-resonant" conditions for reflection from the cladding capillary wall, only the minor semi-axis a of the cladding capillary ellipse will change in the calculations. Therefore, the calculation was performed at a wavelength of $\lambda$ = 1.55 μm for a hollow- core fiber with an air- core diameter of $D_{core}$ = 38μm and a cladding capillary wall thickness of 490 nm (Fig. 1). This wall thickness of the capillary cladding corresponds to the first (longest wavelength) transmission band of the hollow- core fiber. In all subsequent calculations, the semi major axis of the ellipse b will be equal to 20 μm (Fig. 1). Figure 2 shows that the dependence of leakage losses on the a/b value has a pronounced minimum at a/b = 0.6. In this case, the distribution of the radial component of the Poynting vector of the fundamental air- core mode has an oscillating character with negative $P_r$ values. In addition, the $P_r$ distribution is symmetrical relative to the point on the cladding capillary wall closest to the fiber axis. In this area of the capillary wall, the oscillations reach their maximum values (Fig. 2, insert with blue curve). Oscillations of the Pr values occur along the entire capillary wall up to the point of attachment of the cladding capillary to the supporting tube (coordinate 0 on the abscissa axis for insertion with the blue curve in Fig .2). The rotation of the transverse component of the Poynting vector of the air-core mode is determined by the complex shape of the core boundary of the hollow - core fiber, its "negative curvature". In this case, the leakage loss of the air-core mode is determined by both the radial and azimuthal components of the Poynting vector (see Supplementary):

$$\mathrm{Im}(\beta) = \oiint \left(P_r\vec{r} + P_\varphi\vec{\varphi}\right)\vec{n}\,dS / \left(2VP_z\right),$$

where Im(β) is the imaginary part of the propagation constant of the air core mode, $\vec{n}$ is the normal to the surface of the core boundary of the hollow - core fiber, and $V$ is some small volume of the fiber including the core boundary. The presence of negative values of $P_r$ along the boundary of the cladding capillary indicates that vortices of air- core mode energy flow arise in its wall and near it, which generate reverse radial energy flows. Some asymmetry in the distribution of $P_r$ in the insets to Fig. 1 is due to the fact that the process of outflow of the air core mode energy through individual cladding capillaries is associated with the orientation of the electric field vector of the mode relative to the surface of the capillary. This issue requires additional study in the future. It is worth noting that the interaction of the air core mode with different polarization states with the boundaries of the cladding capillaries does not affect the overall level of losses.

Let us consider the distribution of the radial $P_r$ component of the Poynting vector for the fundamental air-core mode in the region of the a/b parameter in which an increase in leakage losses seems counter intuitive from the point of view of the ARROW model, since increasing the value of the minor semi-axis we increasingly overlap the space between the cladding capillaries. The capillary wall thickness and the diameter of the air- core remain the same. With the values of the parameter a/b = 0.85, the leakage loss of the air core fundamental mode is about two orders of magnitude greater than in the case discussed above (Fig.2). The distribution of the radial component of the Poynting vector has a qualitatively different character, although with a region of reverse energy flow (Fig. 2, insert with red curve). In addition, the area of vortices formation of the air- core mode energy flow is limited and near the point of attachment of the capillary to the supporting tube there are only positive values of the mode energy flow (coordinate 0 on the abscissa axis for insertion with the red curve in Fig. 2).

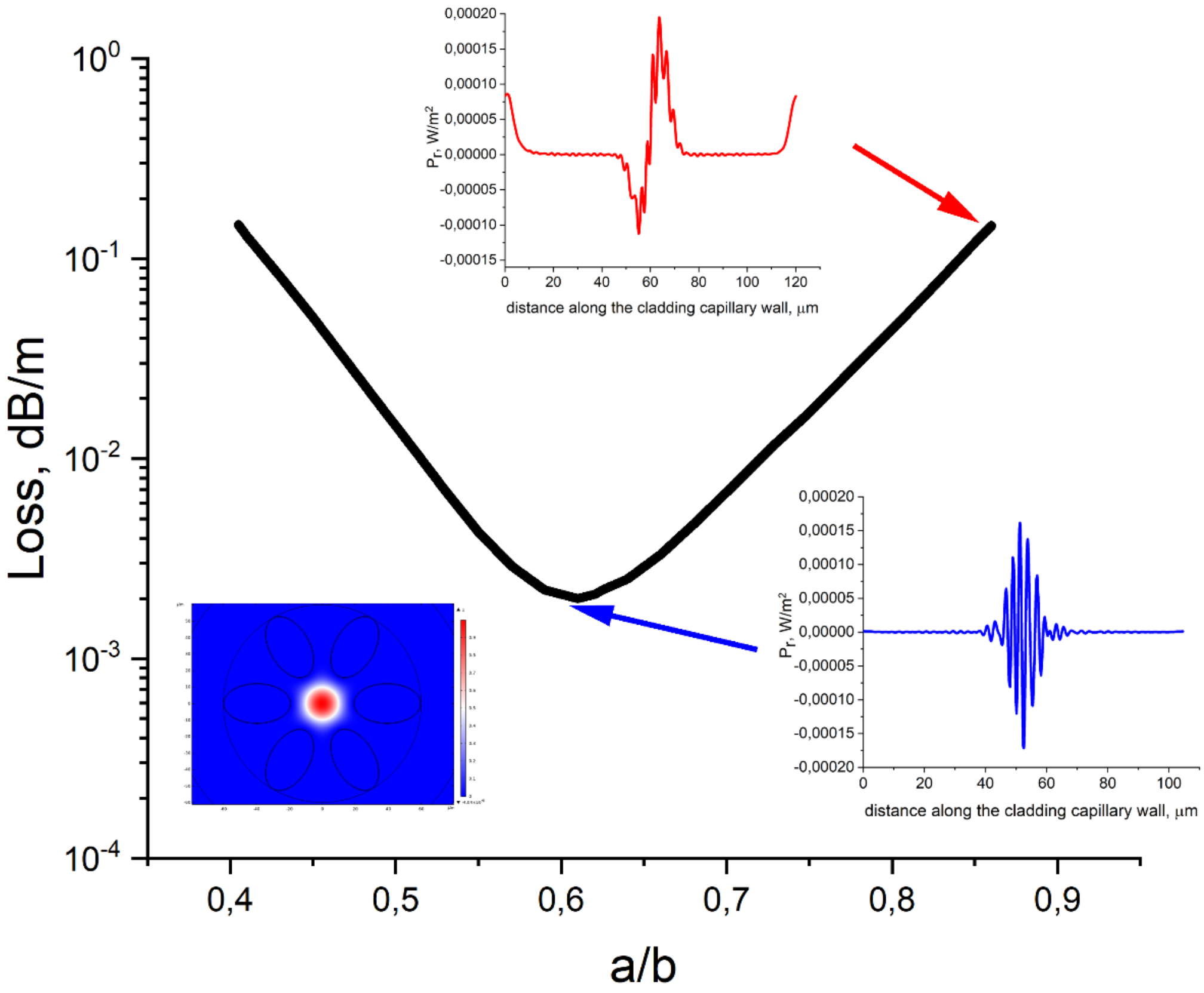

Figure 2: Losses for the fundamental air- core mode of a hollow- core silica glass fiber with six elliptical capillaries in the cladding (color insert), depending on the ratio of the minor to major axis of the ellipse a/b at λ = 1.55 µm. The other two inserts show the distributions of the radial component of the Poynting vector of the fundamental air- core mode for two values a/b = 0.6 (minimum loss, blue curve) and a/b = 0.85 (maximum loss, red curve) along the outer boundary of the cladding capillary.

It is obvious that the increase in losses for the values of the parameter $a/b < 0.6$ is associated with the fact that the distance between the cladding capillaries increases, which leads to an increase in the radial component of the energy flow of the air core mode between the capillaries.

The distributions of the $P_r$ components shown in the inserts in Fig. 2 indicate that there is an optimal distribution of vortices of the energy flow in the cladding capillary wall, which ensures a minimum of leakage losses in the hollow- core fiber.

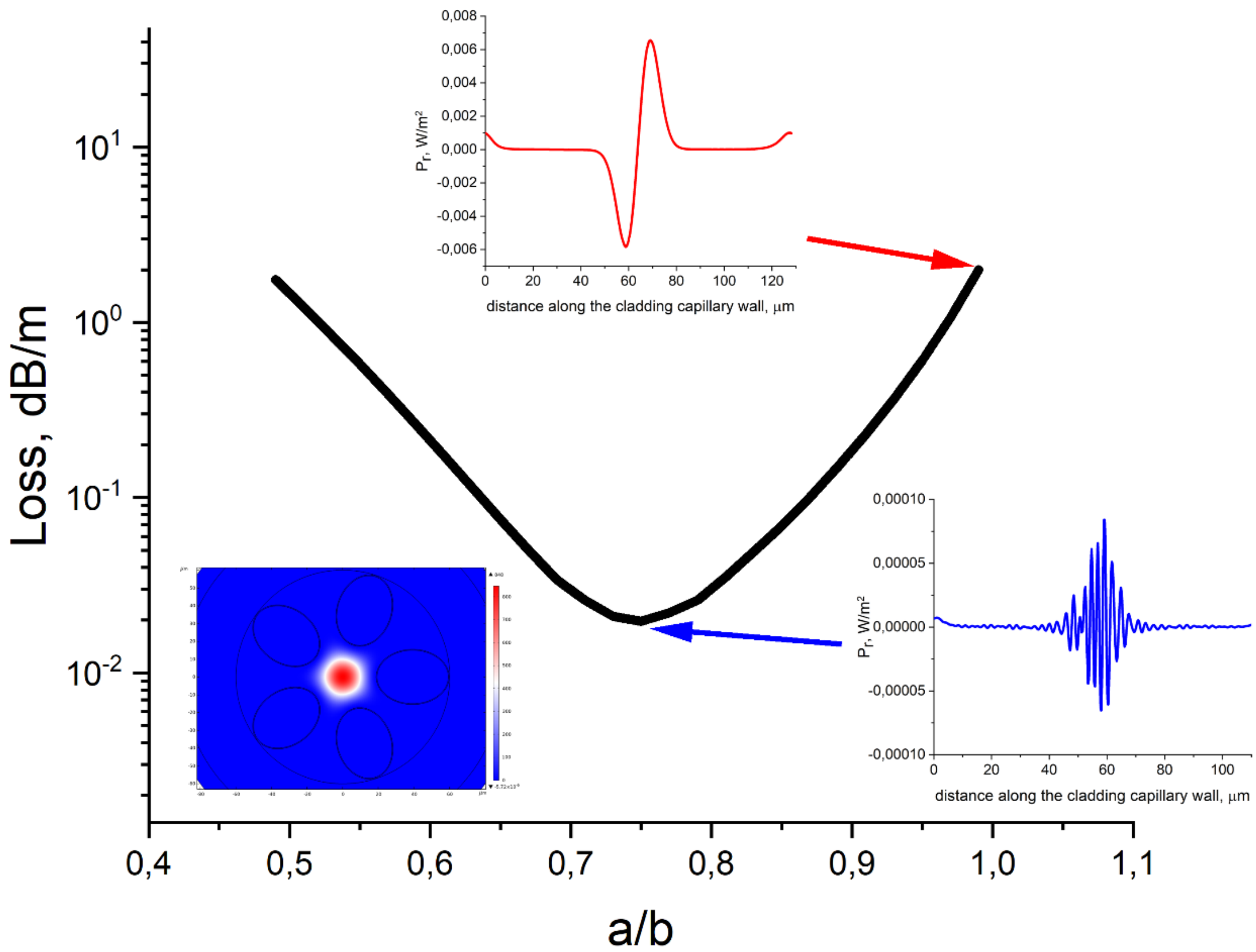

Figure 3: Losses for the fundamental air- core mode of a hollow- core silica glass fiber with five elliptical capillaries in the cladding (color insert), depending on the ratio of the minor to major axis of the ellipse a/b at λ = 1.55 µm. The other two inserts show the distributions of the radial component of the Poynting vector of the fundamental air- core mode for two values a/b = 0.75 (minimum loss, blue curve) and a/b = 1 (maximum loss, red curve) along the outer boundary of the cladding capillary.

To clarify the statement that the distribution of vortices of the energy flow and, accordingly, the reverse energy flows of the air- core mode in the cladding capillary wall directly affect its losses, let us consider another example of a hollow- core fiber.

If we calculate the dependence of losses on the a/b parameter for a hollow- core fiber with five capillaries in the cladding, we can also obtain an optimum that shifts to large values of a/b (Fig. 3). We will also obtain distributions of the radial component of the Poynting vector for the fundamental air- core mode at the outer boundary of the cladding capillary at maximum and minimum losses, which qualitatively coincide with similar distributions in Fig. 2. The $P_r$ distribution oscillates at minimum losses, thereby confirming the vortex nature of the behavior of the air - core mode energy flow in the transverse direction.

At the point of maximum loss at a/b = 1, we get a conventional negative curvature hollow- core fiber with a fairly high loss level despite the fact that the distance between the cladding capillaries is much smaller than in the case of the parameter value a/b = 0.75 at the point of minimum loss (Fig. 3). It is known that the level of air- core mode loss can be reduced by introducing an additional capillary into the cavity of the cladding capillary. How will the distribution of the radial component of the Poynting vector of the air- core mode at the boundary of the cladding capillary change? This question is answered in Fig. 4, where the $P_r$ distributions are shown at the boundaries of the large and small cladding capillaries. The diameter of the small capillary is two times smaller than the diameter of the large one, and its wall thickness is the same. When nested capillaries are added to the cladding, losses in a hollow- core fiber with five capillaries in a cladding with a ratio of a/b = 1 (Fig. 3) became equal to $1.2e-4$ dB/m, that is, they decreased by two orders of magnitude compared to the minimum loss value in Fig. 3. At the same time, there was a qualitative change in the distribution of the radial component of the Poynting vector of the fundamental air- core mode at the boundary of the large cladding capillary (Fig. 4(left)) compared to the similar distribution shown in the inset to Fig. 3 in red. This distribution became similar to the distribution of $P_r$ for small losses in Fig. 3 (blue inset) or in Fig. 2(blue inset); this means that there are vortex movements in the cladding capillary wall and reverse energy flows in the radial direction. A similar distribution of $P_r$ appears in the walls of the small cladding capillaries (nested capillaries) (Fig. 4(right)). From this, it can be concluded that the appearance of vortices in the energy flow of the air- core mode in the cladding capillary walls should lead to a decrease in losses in the hollow- core fiber.

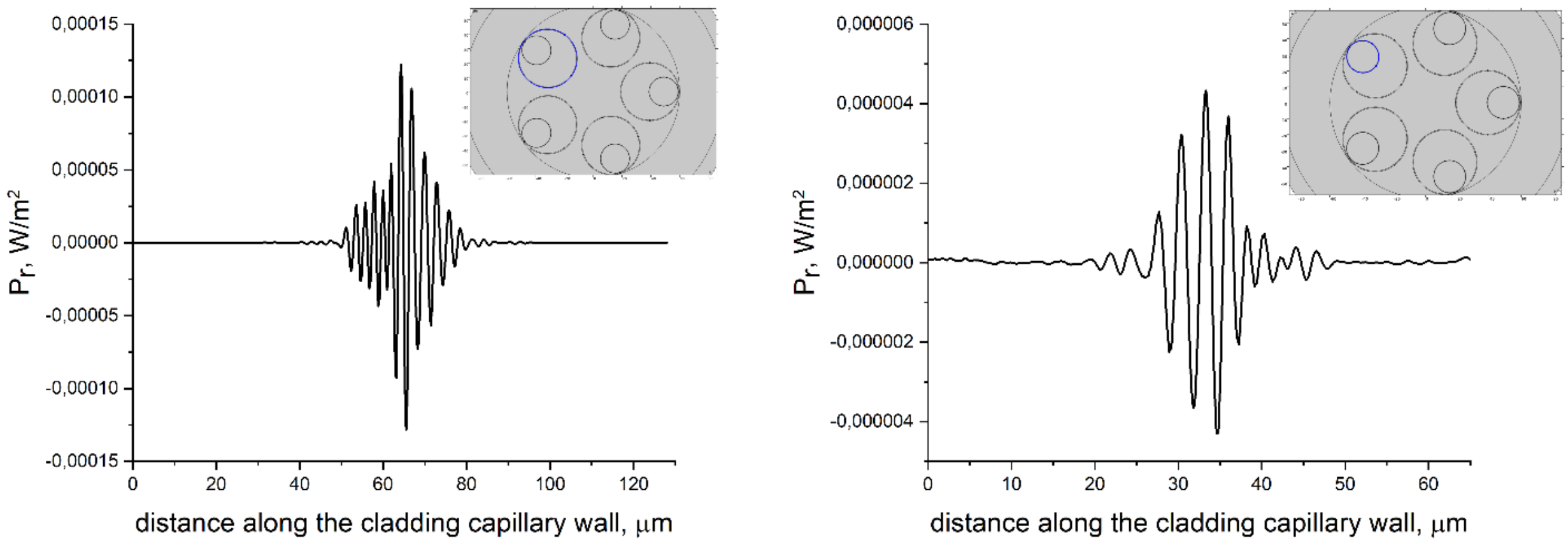

Figure 4: (left) distribution of the radial component of the Poynting vector $P_r$ along the big cladding capillary wall (the boundary of the cladding capillary wall is highlighted in blue on the inset) for the nested hollow- core fiber (inset) for the value of the ratio a/b = 1 in Fig. 3; (right) distribution of the radial component of the Poynting vector $P_r$ along the small cladding capillary wall (the boundary of the small cladding capillary wall is highlighted in blue on the inset) for the nested hollow- core fiber (inset) for the value of the ratio a/b = 1 in Fig. 3.

To conclude this section, we present a figure where the dependences of leakage losses on the parameter of a/b are plotted sequentially for three types of hollow – core fibers. From the Fig. 5. it is seen that the minima of all three loss curves lie practically on the same straight line.

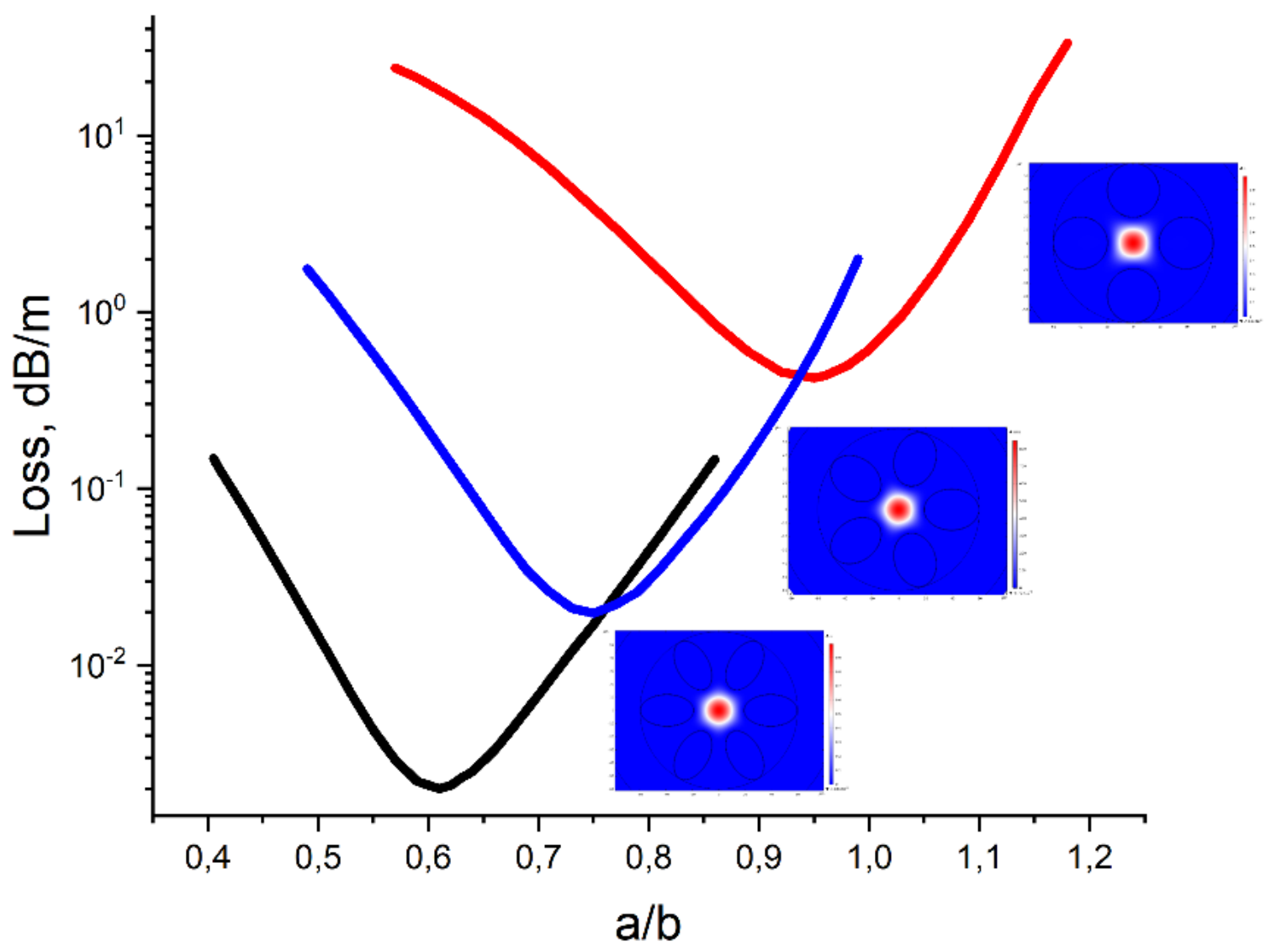

Fig.5. Dependence of losses on the parameter $a / b$ for three hollow – core fibers, two of which were considered in this section and the third has four cladding capillaries and the same $D_{core}$ value and cladding capillary wall thickness.

## 2.2 The behavior of the zero lines of the electric field components and the zero lines of the Poynting vector of the fundamental air- core mode.

Obviously, all the hollow- core fibers discussed in the previous section (regardless of the shape of their cladding elements) transmit light in an antiresonance regime according to the ARROW model. Since the wall thickness of the cladding capillary in all cases is $d$ = 490 nm, according to the formula $\frac{2\pi d}{\lambda}\sqrt{n^2-1}=(m+1/2)\pi$, where $n$ is a refractive index of silica glass at wavelength of $\lambda$ = 1.55 µm, the phase incursion of the air- core mode field in the cladding capillary wall corresponds to the first radiation transmission band at $m$ = 0 for all considered hollow- core fibers. This means that all components of the fundamental air- core vector mode should be well reflected from the cladding capillary wall. As can be seen from Fig. 6, the phase of the component $E_\varphi$ of the electric field of the fundamental air- core mode has a jump when passing through the walls of the cladding capillaries equal to π. This occurs both in the case of minimum losses and in the case of maximum losses. Similar distributions of the phase of the transverse components of the electric and magnetic fields of the fundamental air- core mode can be obtained. That is, the phase conditions for the transverse components of the fundamental air- core mode fields are the same when they reflect from the cladding capillary wall. These conditions do not quite agree with the ARROW model, we will discuss this below. What is different for the case of small and large losses of the fundamental air- core mode shown in Fig. 2? As can be seen from Fig. 6, the difference lies in the distributions of zero values curves $P_r(r,\varphi) = P_\varphi(r,\varphi) = 0$ the intersections of which, according to the rules of singular optics, form centers for the vortex motions of the air- core mode energy.

The phase jumps of π magnitude when passing through the cladding capillary wall can also be explained based on the principles of singular optics. It is known that in a scalar electric field, when curves Re(E) = Im(E) = 0 are superimposed the so-called phase edge dislocation occurs. It is precisely when passing through the phase edge dislocation that the phase changes abruptly, the value of which is equal to π. The air- core vector modes of the hollow- core fibers are leaky and have corresponding propagation constants $\beta = Re(\beta) + Im(\beta)$, where $\beta = 2\pi n_{eff}/\lambda$ and $n_{eff}$ is an effective refractive index of the mode. This means that the fields components of these modes have the form $E\alpha = Re(E_\alpha) + iIm(E_\alpha) = |E_\alpha|e^{i\phi}$ , where $\alpha = (r,\varphi,z)$ and $\phi$ is the phase of electric or magnetic field component. When the amplitude is zero, there is a phase uncertainty for this field component, which leads to the formation of phase edge dislocation for the field component in the cladding capillary wall.

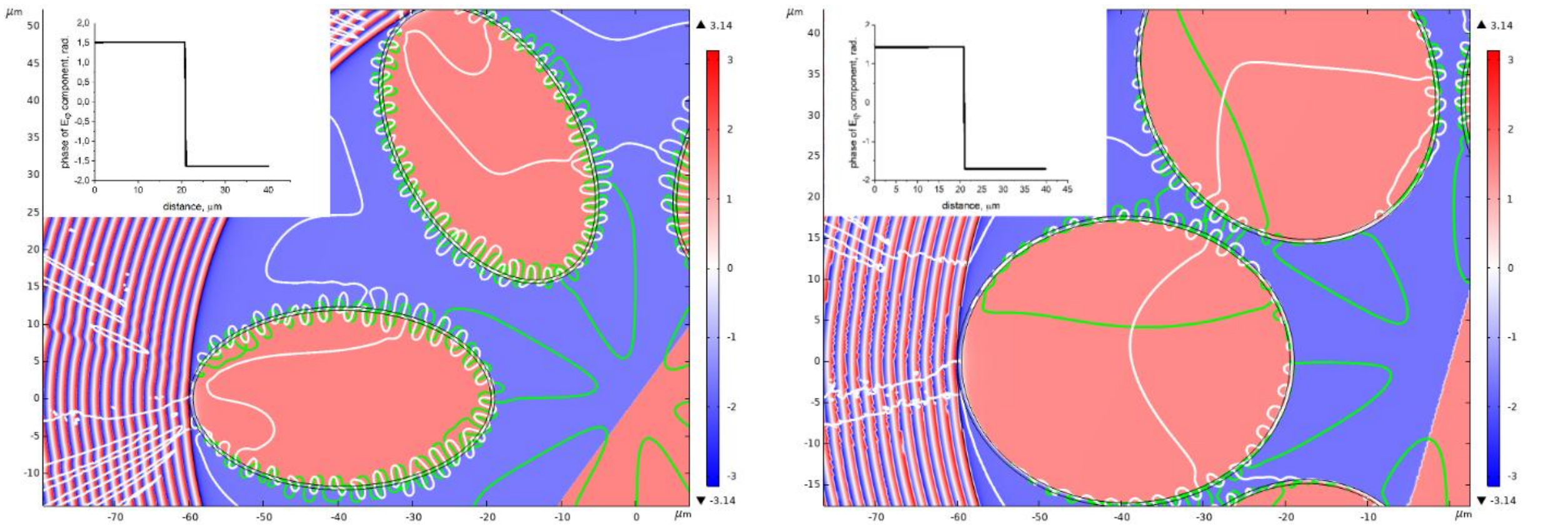

Figure 6: (left) lines of zero values for $P_r(r,\varphi) = 0$ (green) and $P_\varphi(r,\varphi) = 0$ (white) for the fundamental air- core mode shown in Fig. 2 (inset) at the point of minimum losses a/b = 0.6, their intersections in the capillary wall lead to the appearance of vortices of the air- core mode energy flow in the cross- section; the color distribution is the phase for the component $E_\varphi$ of the electric field of this mode, and the inset shows a phase jump equal to $\pi$ that occurs when passing through the cladding capillary wall; (right) the same lines and distributions for the high loss case in Fig. 2 at a/b = 0.85.

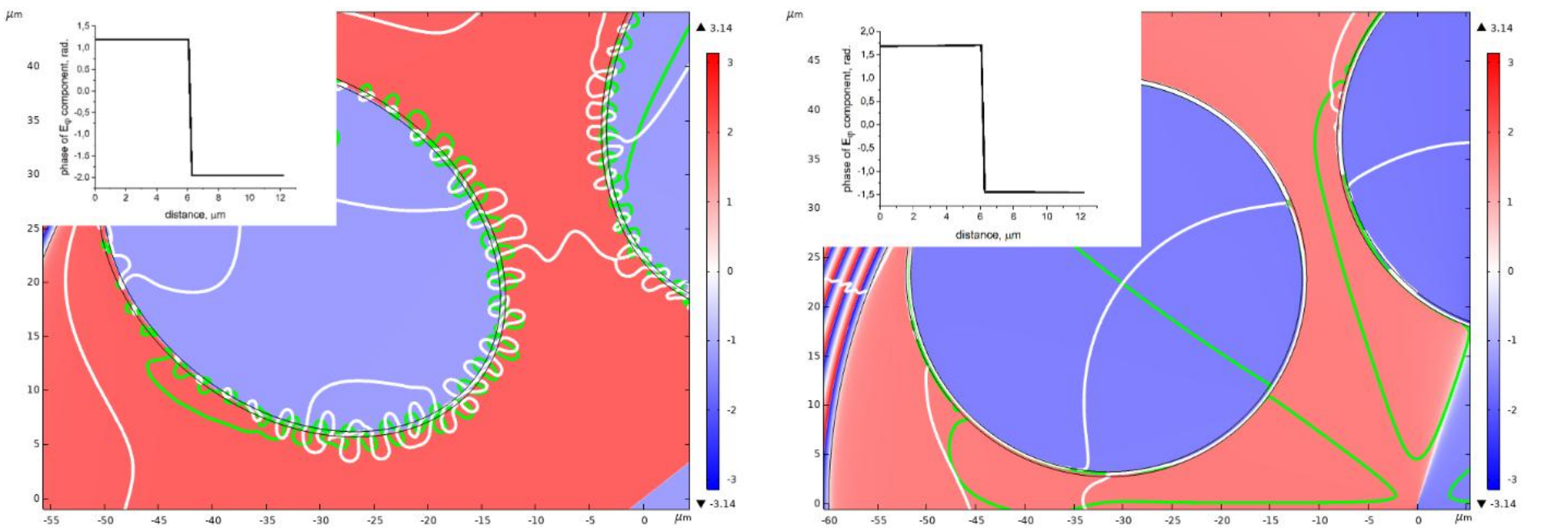

Figure 7: (left) lines of zero values for $P_r(r,\varphi) = 0$ (green) and $P_\varphi(r,\varphi) = 0$ (white) for the fundamental air- core mode shown in Fig. 3 (inset) at the point of minimum losses a/b = 0.75, their intersections in the capillary wall lead to the appearance of vortices of the air- core mode energy flow in the cross- section; the color distribution is the phase for the component $E\varphi$ of the electric field of this mode, and the inset shows a phase jump equal to $\pi$ that occurs when passing through the cladding capillary wall; (right) the same lines and distributions for the high loss case in Fig. 3 at a/b = 1; in this case, there are no vortices of the fundamental air- core mode energy flow in the capillary wall.

Figure 7 shows the phase edge dislocation and the corresponding phase jump of π for the azimuthal component of the electric field of the fundamental air - core mode and zero values curves $P_r(r, \varphi) = P_\varphi(r, \varphi) = 0$ for a hollow - core fiber with five capillaries in the cladding in the case of maximum and minimum losses (Fig. 3).

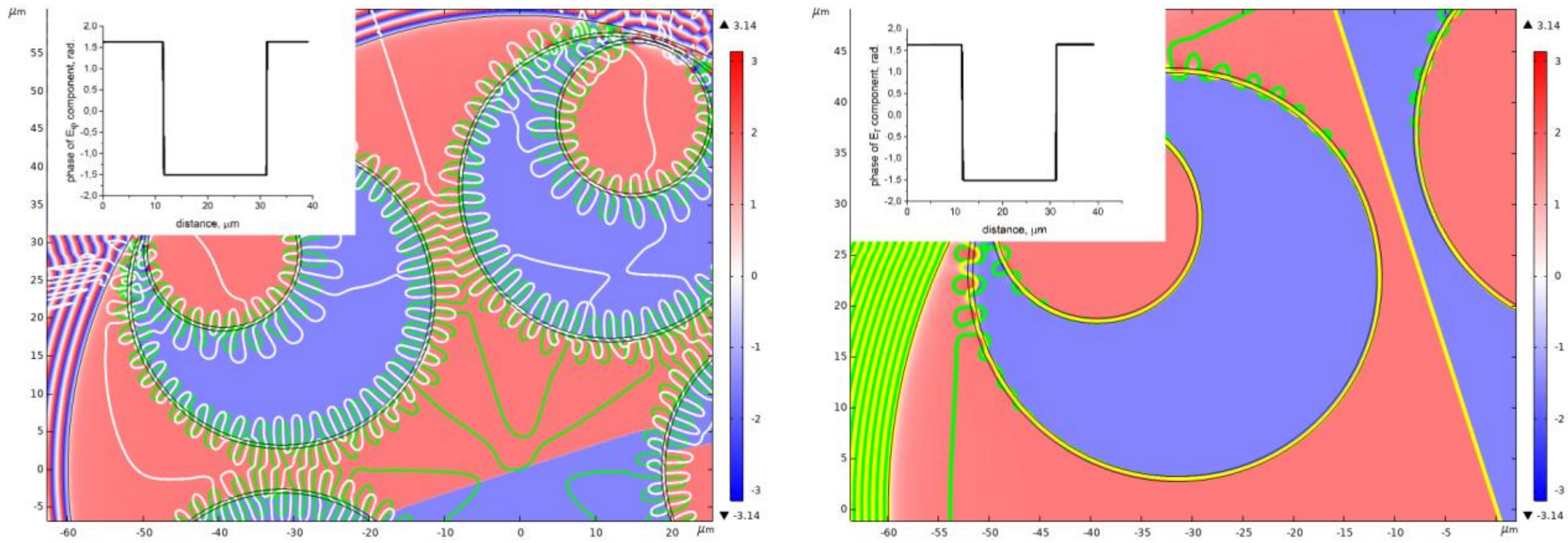

Figure 8: (left) zero values curves for $P_r(r,\phi) = 0$ (green) and $P_\varphi(r,\phi) = 0$ (white) for the fundamental air- core mode of nested hollow- core fiber described in the text, their intersections lead to a dense distribution of vortices of the air- core mode energy flow in the wall of both the large and small cladding capillaries (Fig.4); the color distribution is the phase for the component $E_\varphi$ of the electric field of this mode, and the inset shows a phase jump equal to π that occurs when passing through the walls of both cladding capillaries. (right) phase distribution for the radial component of the electric field $E_r$ for the same mode; phase edge dislocations for $E_r$ are observed due to overlapping curves $Re(E_r(r,\varphi)) = 0$ (green) and $Im(E_r(r,\varphi)) = 0$ (yellow) in the cladding capillary wall.

The appearance of phase edge dislocations is very well manifested in nested hollow- core fiber, which was described above (Fig. 4). Figure 8 (left) shows the phase distributions of the azimuthal component of the electric field and the zero values curves $P_r(r,\varphi) = P_\varphi(r,\varphi) = 0$ for the fundamental air- core mode. As can be seen from the distribution of zero values curves the density of vortices of the transverse component of the Poynting vector increases strongly in the cladding capillary walls, and in addition, the centers of vortices of the transverse energy flow are also observed in the space between the cladding capillaries, which leads to additional blocking of the flowing mode energy in the radial direction.

In order to show that a phase jump of π is characteristic of all transverse components of the electric and magnetic fields of the air- core mode, the phase distribution for component $E_r$ was calculated, from which similar abrupt phase changes are clearly visible (Fig. 8 (right)). In addition, Fig. 8(right) shows curves $Re(E_r(r,\varphi)) = 0$ and $Im(E_r(r,\varphi)) = 0$ that overlap each other in the capillary wall, leading to phase edge dislocations for the selected component of the fundamental air- core mode field and the corresponding phase jump. The curves $Re(E_r(r,\varphi)) = 0$ and $Im(E_r(r,\varphi)) = 0$ become oscillating only closer to the point of attachment of the capillary to the supporting tube, which leads to incomplete overlap of these curves. Similar phase distributions can be demonstrated for the transverse components of the magnetic field of the fundamental air- core mode. Therefore, all transverse components of the air- core mode fields oscillate in antiphase outside and inside the cladding capillaries in the antiresonant light reflection regime. The main differences that arise when changing the shape of the cladding capillary are associated with a change in the distribution of singularities of the transverse component of the Poynting vector of the air- core mode in the walls of the cladding capillaries and, as a consequence, to different distributions of direct and reverse radial energy flows of the mode.

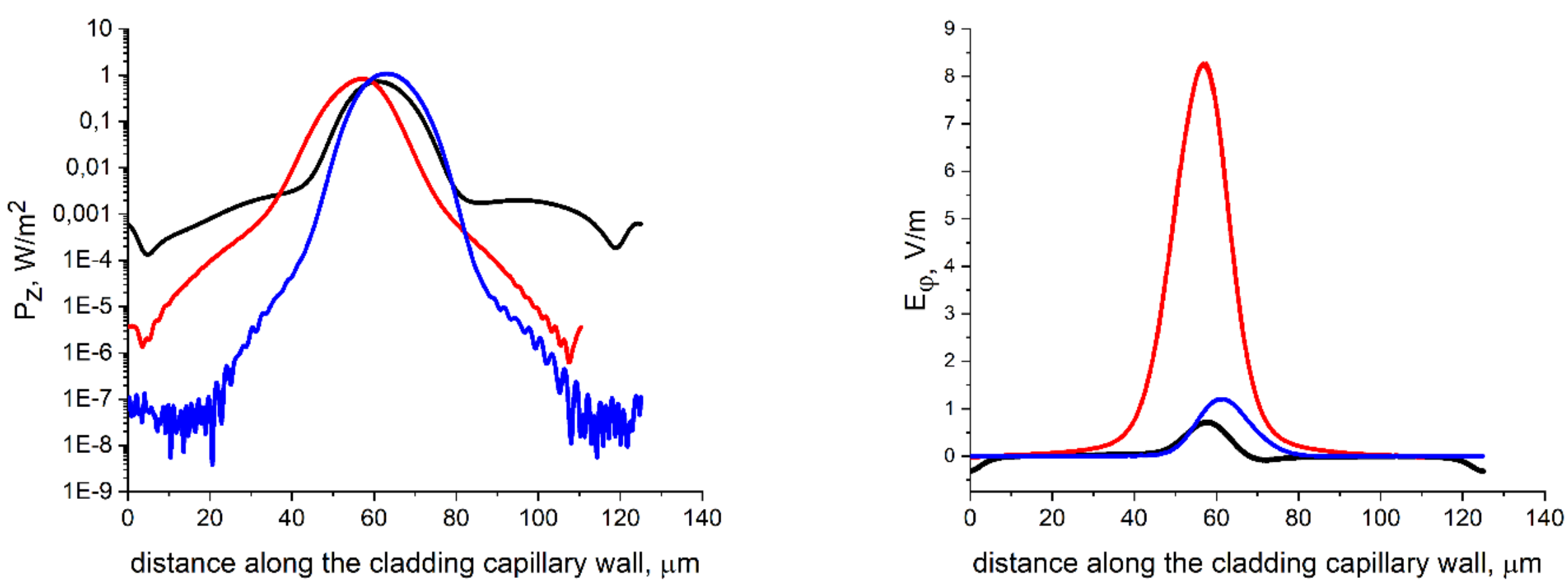

Figure 9: (left) distributions of $P_z(r,\varphi)$ along the cladding capillary wall for the fundamental air- core mode of hollow core fibers shown in Fig. 7(left) (red), Fig. 7(right) (black), Fig. 8 (big cladding capillary) (blue) ; (right) the same distributions for $E_\varphi$.

The distributions of the electromagnetic field components of the air- core mode along the cladding capillary wall directly affect the distributions of the components of the Poynting vector of the mode. In the case of the radial component of the Poynting vector, there is a maximum of the envelope $P_r$ in the area of the cladding capillary wall closest to the fiber axis, and then a sharp decrease occurs up to the point of attachment of the cladding capillary to the supporting tube (Fig. 4). Oscillations of $P_r$ values occur under the envelope due to the presence of vortices in the transverse component of the Poynting vector in the capillary wall. The axial component of the $P_z$ mode has a similar distribution, but without negative values (Fig. 9(left)). These distributions are also determined by the distributions of the components of the electric and magnetic fields of the air- core mode (Fig. 9(right)). The components of the air- core mode fields also have maximum values in the region of the capillary wall closest to the fiber axis and then decrease towards the attachment point of the cladding capillary to the supporting tube. The significantly higher value of the azimuthal component of the electric field in Fig. 9(right) (red) is determined, in our opinion, by the shape of the cladding capillary with a value of a/b = 0.75 (Fig. 7(left)). Therefore, the phase characteristics of the electric and magnetic fields of the air- core mode turn out to be the same in the case of antiresonance due to the presence of phase edge dislocations in the cladding capillary walls of the hollow- core fibers. The differences are manifested in the behavior of the transverse components of the Poynting vector for different cladding capillary shapes or when a nested capillary is introduced. Despite the qualitatively similar distributions of the $P_z$ and $P_r$ components along the cladding capillary wall, in the case of the radial component of the Poynting vector, we have its oscillatory character with periodic values of $Pr < 0$, which leads to a difference in losses for hollow- core optical fibers.

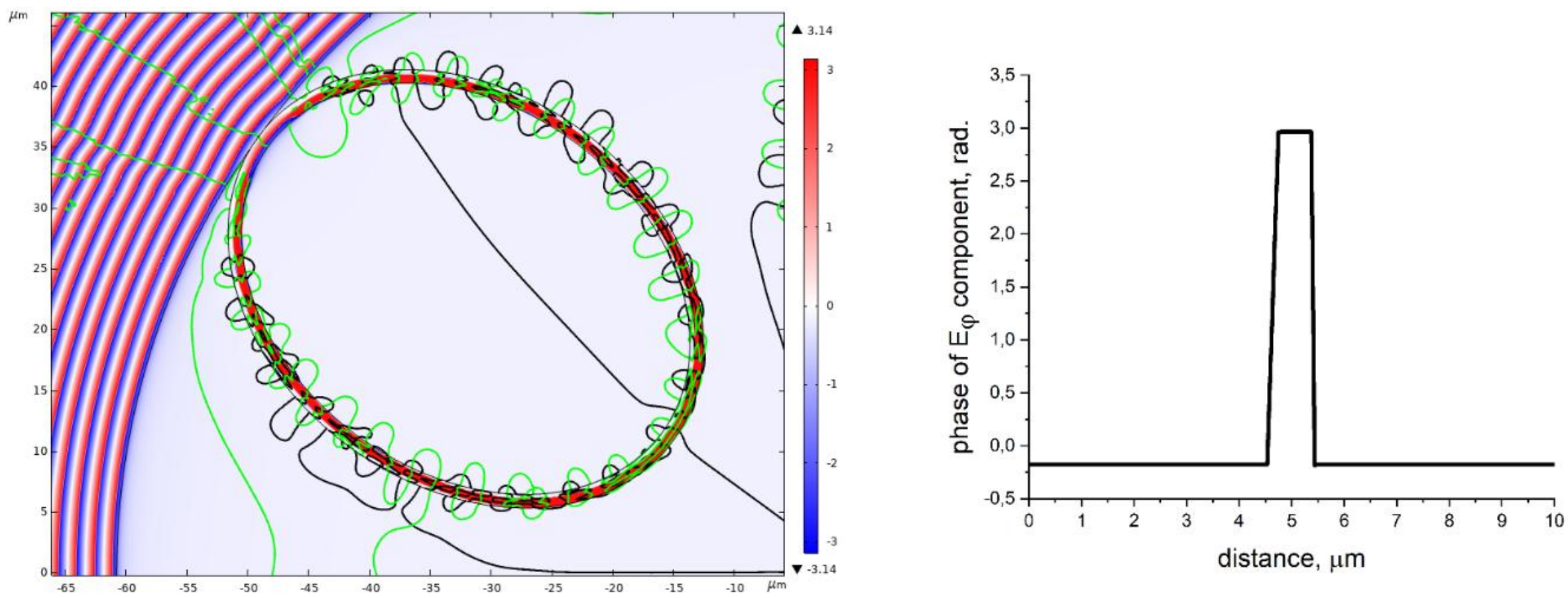

Figure 10: (left) lines of zero values for $P_r(r,\varphi) = 0$ (green) and $P_\varphi(r,\varphi) = 0$ (black) for the fundamental air- core mode in the second transmission band for the hollow – core fiber shown in the inset to Fig. 3, their intersections in the capillary wall lead to the appearance of vortices of the air- core mode energy flow in the cross- section; the color distribution is the phase for the component $E\varphi$ of the electric field of this mode; (right) a phase jump equal to $\pi$ that occurs when passing through the cladding capillary wall, in the second transmission band two phase dislocations are formed, therefore the phase jump for the electric field component occurs between them.

To conclude this section, we will demonstrate the presence of phase dislocations for hollow – core fibers in the subsequent transmission bands. Let us consider the phase distribution of the electric field component of the fundamental air - core mode for the fiber shown in the inset to Fig.3 in the second transmission band. For this purpose, we will leave the working wavelength the same as before, 1.55 μm, but change the wall thickness of the cladding capillaries 1.12 μm. From Fig. 10(left) it is seen that the behavior of the curves and  has not changed qualitatively compared to the previously considered cases and therefore the formation of Poynting vector vortices of the mode with reverse energy flows in the radial direction occurs. In addition, two phase edge dislocations are formed in the capillary wall for the transverse components of the air - core mode fields with a corresponding phase jump π between them (Fig.10(right)), since this is the second transmission band for the hollow - core fiber.

## 3. Discussion

Two main models have been established to describe strong light localization in hollow-core optical fibers: the ARROW (Antiresonant Reflecting Optical Waveguide) model and the inhibited coupling model. Notably, both frameworks overlook the influence of the negative curvature of the core boundary in accounting for the exceptionally low optical losses observed in such fibers.

In the ARROW model, the analysis is limited to the optical thickness of the cladding capillary walls and the corresponding phase conditions that arise from the interference of rays reflected at the inner and outer boundaries of the cladding. By adjusting the phase shift of the beam reflected from the inner interface, it is possible to achieve either maximal reflection (antiresonance) or maximal transmission (resonance) of radiation.

Our results reveal that in the regime of antiresonance (or minimal loss), a distinctive phase relationship consistently exists in all configurations of cladding capillaries, irrespective of their number or geometrical arrangement within the fiber. Specifically, each component of the electromagnetic fields associated with the air-core vector mode exhibits a phase discontinuity of $\pi$ as it traverses the capillary wall. This phase jump can be naturally interpreted within the framework of singular optics as the manifestation of phase edge dislocations localized within the capillary walls. The air-core mode fields on both sides of the cladding capillary wall oscillate in anti-phase for different capillary shapes provided their wall thickness is the same. Therefore, the strong confinement of light in hollow-core fibers with low transmission losses cannot be fully explained by the usual arguments related to the phase shift in the anti-resonant transmission regime since losses for different shapes of the cladding capillary can vary significantly in this case.

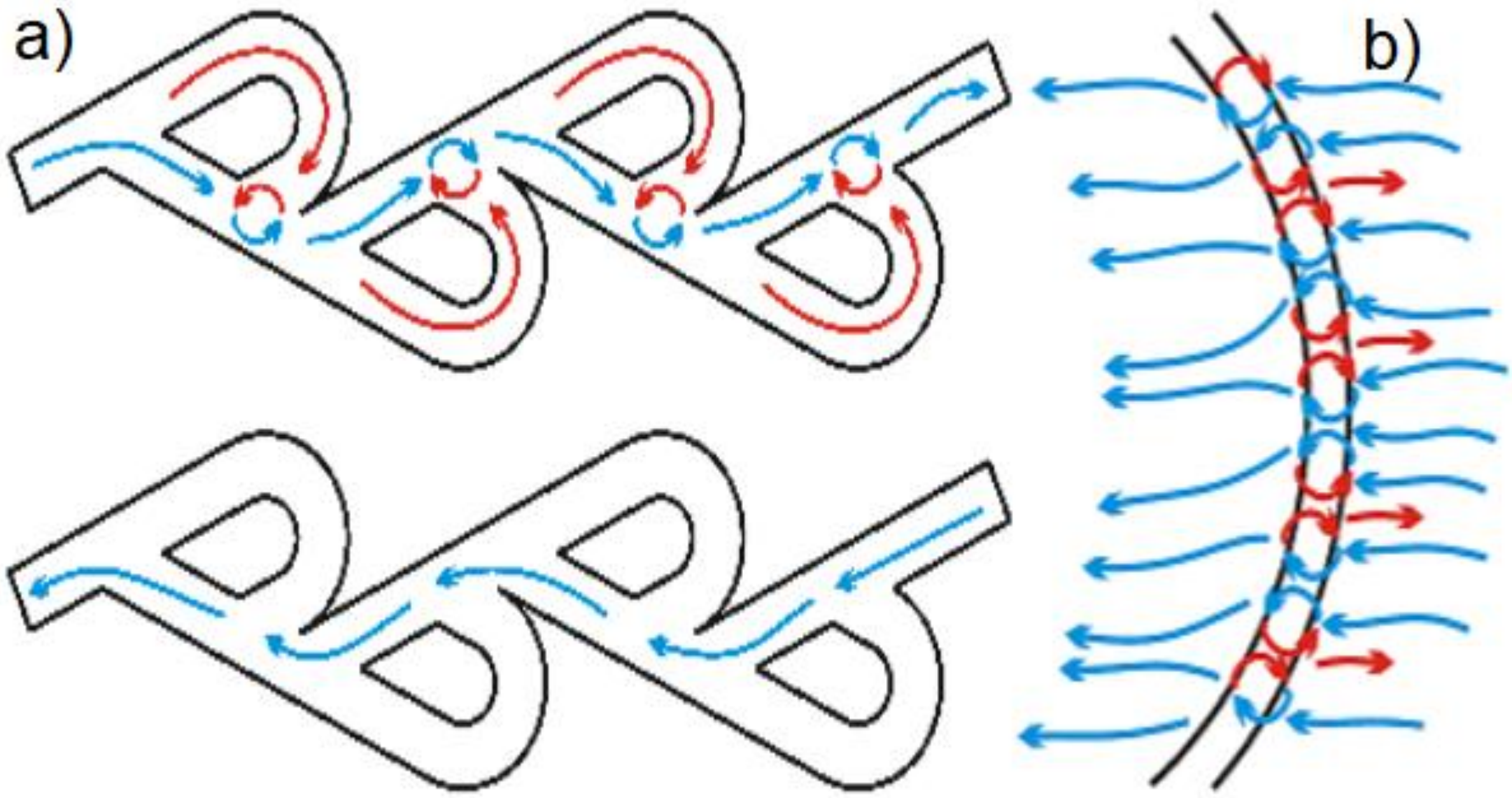

Figure 11. (a) Schematic illustration of the analogy between the fluid flow in a Tesla valve and the energy flows in HCF. The flow in a Tesla valve in one direction is slowing down due to the merging of forward and reverse flows (top), while in the opposite direction the flow of liquid moves without resistance due to the absence of vortices (bottom). (b) depicts the emergence of direct and reverse energy flows of air-core mode in the cladding capillary wall of the hollow core fiber due to the formation of vortices of the transverse component of the Poynting vector.

The approach associated with calculating the overlap integrals between the air-core and cladding modes and searching for its minimum (inhibited coupling model) does not provide a deeper insight into the physical mechanisms. In particular, it does not explain the nature of such reduced losses in relatively simple waveguide structures as hollow-core fibers with cladding consisting of capillaries.

An interesting approach has recently been introduced based on consideration of the azimuthal confinement in hollow-core fibers [27]. This theory states that the part of the cladding capillary wall closest to the axis of the hollow-core fiber reflects light according to the ARROW model. However, for capillaries of any shape with low air-core mode losses, there are phase edge dislocations in the capillary wall, and the phase in the cladding capillary wall changes abruptly. The components of the electric and magnetic fields of the air-core mode decrease in the wall of the cladding capillary along with the energy fluxes of the air-core mode, and the rate of decrease of their values depends very much on the shape of the cladding capillary at a given wavelength. As our calculations in Section 2.1 demonstrated, the radiation reflection from the cladding capillaries strongly depends on their shape ('negative curvature') when all other parameters are the same. The most important observation is that the circle cladding capillaries, which cover the space between them as much as possible, do not provide for low losses compared to ellipsoidal capillaries. This is due to the behavior of the transverse energy fluxes of the air-core mode, which is an additional important factor that occurs when radiation is reflected in such hollow-core fibers.

To conclude, we would like to mention a fascinating analogy between the effect of the reverse radial energy flows of the air-core modes on losses in hollow-core fibers and the hydrodynamical phenomenon related to the operation of the passive fluid guiding device invented in 1920 by Nikola Tesla - Tesla valve. Valves are structures with a higher pressure drop for fluid flow in the reverse direction than in the forward direction. Tesla valve is a series of interconnected, asymmetric, tear-shaped loops (islands and bends) that allow it to move fluid in a single direction without any moving parts. This device generates a substantial resistance difference between two flow directions.

When fluid flows in the desirable direction, it moves smoothly through the channels with minimal resistance. However, for the flow in the reverse direction, the geometry of the valve creates vortices disrupting the flow and increasing the resistance to the backward leakage. The operating principle of Tesla valve is now used in different applications [28, 29]. In our case, when vortices of the transverse component of the Poynting vector of the air-core mode appear, direct and reverse flows of the mode energy arise near the centre of these vortices, which, by analogy with the flows of liquid in the Tesla valve, mix and slow down the flow of the air- core mode energy in the radial direction (Fig. 11). The effect of the air-core mode energy flow reduction in the radial direction in the presence of vortices resulting from the interaction of direct and reverse mode energy flows and the corresponding decrease in the local velocity of the air-core mode energy in the cladding capillary wall was discussed in [30]. The level of leakage losses is determined by alternating the radial flow of the air- core mode energy in the places of vortex formation. Since the amount of outgoing mode energy in the radial direction is always greater than the amount of energy flowing in the opposite direction, the direct and reverse energy flows are always asymmetrical, as in the case of the Tesla valve (Fig. 11). The distribution of the alternating radial energy flux of the air-core mode in the cladding capillary wall is determined, as shown above, by the number of cladding capillaries and their shape, all other things being equal. Note that the maximum overlap of the space between the cladding capillaries does not provide an optimum for losses. In order to confirm this fact, we can consider the recently proposed hollow – core fiber [31] in which the total losses reach the level of 0.1 dB/km. In this case, it is necessary to use 3 nested cladding capillaries instead of one. To achieve this level of loss, a hollow – core fiber with an air - core diameter of about 30 μm was used at a wavelength of 1.55 μm in the longest wavelength transmission band. It can be shown (Supplementary) that there is no need to use three capillaries inserted into each other in the fiber cladding, but a significant reduction in leakage losses can be achieved by changing the shape of the surface of the cladding capillaries (their "negative" curvature) and by changing the number of cladding capillaries. This occurs due to a change in the distribution of the Poynting vector vortices and, accordingly, the reverse energy flows of the air - core modes in the cladding capillary walls (Supplementary). It should be noted that the phase relationships for the components of the air-core mode fields remain the same ("anti-resonance"), but the decisive role is played by the nonlinear dynamics of these field components associated with the Poynting vector.

The optimum is achieved by an arrangement of vortices in the capillary wall that provides the maximum of the reverse energy flows of the air-core mode. As stated in the Introduction, another cross-field connection is the reverse radial and azimuthal energy flows that has already been experimentally demonstrated in OAM beams. We anticipate that our work will stimulate further analysis of the impact of the reverse energy fluxes in HCFs and the better understanding of the physical mechanisms underlying loss reduction will lead to new fiber designs with superior properties.

# 4. Methods

To describe a mechanism of light localization in hollow – core optical fibres associated with the formation of vortices of the Poynting vector of the air - core mode and calculate losses of the leaky air - core modes of the fibres, we employed a commercial sofware package COMSOL Multiphysics based on finite element method (FEM) accounting for all electric and magnetic components of the air - core mode fields and their complex propagation constants. The complex propagation constants and distribution of the fields of the fundamental leaky air - core mode have been found by solving the eigenvalue problem for the vector wave equation for two hollow - core fibres. The waveguide losses have been calculated through the imaginary parts of the propagation constant of the fundamental air - core mode by the above method for different geometrical parameters of the fibres at wavelengths of 1.55 μm corresponding to the first longest wavelength transmission band and at thicknesses of cladding capillaries corresponding to the second transmission band of the hollow – core fibers under consideration. Further, using the calculated values of the air - core mode fields, we computed the distribution of radial $P_r$ and azimuthal $P_\varphi$ projections of the transverse component of the Poynting vector of the fundamental air – core mode. The positions of the optical vortices in the cross – section of the hollow -core fibres ($r_0$, $\varphi_0$) were determined from the equation $P_r$ ($r_0$, $\varphi_0$) = $P_\varphi$ ($r_0$, $\varphi_0$) = 0. The conditions $Re(E_\alpha (r, \varphi)) = Im(E_\alpha (r, \varphi)) = 0$ ( $\alpha = r, \varphi$) defined the singular lines called phase edge dislocations in the cross section of the hollow - core fibres. Optical vortices and phase edge dislocations have been detected in the regions and individual points, where the lines of zero values described above overlap or intersect. The analysis of the vortex distributions and loss dependencies allows us to identify us the connection between the structure of the radial projection of transverse component of the Poynting vector of the fundamental air – core mode and the minimum loss values in the hollow – core fibres.

# Acknowledgement

The authors would like to thank Alexey Kosolapov and Dmitry Komissarov for their help in designing the drawings for this article and useful discussions. SKT acknowledges the support of the EPSRC Programme Grant TRANSNET.

# Data availability

All the data supporting this study are available upon request to the corresponding authors.

## Conflict of interest

The authors declare no competing interests.

## References

[1] Wheeler, W. Apparatus for lighting dwellings or other structures, https://patents.google.com/patent/US247229A/en, US Patent 247,227 (1881).

[2] Poletti, F. Nested antiresonant nodeless hollow core fiber. *Opt. Express* **22**, 23807 – 23828 (2014).

[3] Poletti, F., Wheeler, N., Petrovich, M., Baddela, N., Fokoua, E. N., et al. Towards high-capacity fibre-optic communications at the speed of light in vacuum. *Nat. Photonics* **7**, 279 - 284 (2013).

[4] Fini, J. M., Nicholson, J. W., Windeler, R. S., Monberg, E. M., Meng, L., et al. Low-loss hollow-core fibers with improved single-modedness. *Opt. Express* **21**, 6233 - 6242 (2013).

[5] Poletti, F., Petrovich, M. N., Richardson, D. J. Hollow-core photonic bandgap fibers: technology and applications. *Nanophotonics*, **2**, 315 - 340 (2013).

[6] Ferreira, M. F. S., Rehan, M., Mishra, V., Varshney, S. K., Poletti, F. et al. Roadmap on specialty optical fibers. *JPhysics: Photonics* **7**, 012501 (2025).

[7] Pryamikov, A. D., Biriukov, A. S., Kosolapov, A. F., Plotnichenko, V. G., Semjonov, S. L. & Dianov, E. M. Demonstration of a waveguide regime for a silica hollow - core microstructured optical fiber with a negative curvature of the core boundary in the spectral region > 3.5 μm. *Opt. Express* **19**, 1441 - 1448 (2011).

[8] Chen, Y., Petrovich, M., Fokoua, E. N., Adamu, A., Hassan, M., et. al. In Optical Fiber Communication Conference (OFC) 2024. Optica Publishing Group, 2024 Th4A.8, URL https://opg.optica.org/abstract.cfm?URI=OFC-2024-Th4A.8.

[9] Hong, Y., Almonacil, S., Mardoyan, H., Carrero, C. C., Osuna, S., et .al. Beyond Terabit/s/λ Nonlinearity-Free Transmission Over the Hollow-Core Fiber. *J. Light. Technol*. **43**, 6306 - 6312 (2025).

[10] Mukasa, K., & Takagi, T. Hollow core fiber cable technologies. *Opt. Fiber Technol*. **80**, 103447 (2023).

[11] Tyndall, J. On the action of rays of high refrangibility upon gaseous matter. *Proc. R. Soc. of London* **19,** 114 - 122 (1870).

[12] Yeh, P., Yariv, A., Marom, E. Theory of Bragg fiber. *J. Opt. Soc. Am.* **68**, 1196 – 1201 (1978).

[13] Birks, T., Roberts, P., Russell, P., Atkin, D., Shepherd, T. Full 2-D photonic bandgaps in silica/air structures. *Electron. Lett.* **31**, 1941 (1995).

[14] Russell, P. S. Photonic-Crystal Fibers. *J. Light. Technol*. **24**, 4729 – 4749 (2006).

[15] Roberts, P. J., Couny, F., Sabert, H., Mangan, B. J., Williams, D. P. et. al. Ultimate low loss of hollow-core photonic crystal fibres. *Opt. Express* **13**, 236 - 244 (2005).

[16] Duguay, M. A., Kokubun, Y., Koch, T. L., Pfeiffer, L. Antiresonant reflecting optical waveguides in SiO2-Si multilayer structures. *Appl. Phys. Lett*. **49**, 13 – 15 (1986).

[17] Litchinitser, N. M., Abeeluck, A. K., Headley, C., Eggleton, B. J. Antiresonant reflecting photonic crystal optical waveguides. *Opt. Lett*. **27**, 1592 - 1594 (2002).

[18] Debord, B., Amsanpally, A., Chafer, M., Baz, A., Maurel, M., et. al. Ultralow transmission loss in inhibited-coupling guiding hollow fibers. *Optica* **4**, 209 - 217 (2017).

[19] Kolyadin, N., Kosolapov, a. F., Pryamikov, A. D., Biriukov, A. S., Plotnichenko, V. G., Dianov, E. M. Light transmission in negative curvature hollow core fiber in extremely high material loss region. *Opt. Express* **21,** 9514 - 9519 (2013).

[20] Wang, Y. Y., Wheeler, N. V., Couny, F., Roberts, P. J., Benabid, F. Low loss broadband transmission in hypocycloid-core Kagome hollow-core photonic crystal fiber. *Opt. Lett*. **36**, 669 - 671 (2011).

[21] Belardi, W., Knight, J. C. Effect of core boundary curvature on the confinement losses of hollow antiresonant fibers. *Opt. Express* **21**, 21912 - 21917 (2013).

[22] Belardi, W., Knight, J. C. Hollow antiresonant fibers with reduced attenuation. *Opt. Lett*. **39**, 1853 – 1856 (2014).

[23] Habib, S. M., Bang, O., Bache, M. Low-loss single-mode hollow-core fiber with anisotropic anti-resonant elements. *Opt. Express* **24**, 8429 - 8436 (2016).

[24] Putten, L. D., Fokoua, E. N., Mousavi, S. M. A., Belardi, W., Chaudhuri, S. et. al. Exploring the Effect of the Core Boundary Curvature in Hollow Antiresonant Fibers. *PTL* **29**, 263 - 266 (2017).

[25] Yan, W., Gao, Y., Yuan, Z., Long, X., Chen, Z. et. al. Energy-flow-reversing dynamics in vortex beams: OAM-independent propagation and enhanced resilience. *Optica* **11**, 531 - 541 (2024).

[26] Ghosh, B., Daniel, A., Gorzkowski, B., Lapkiewicz, R. Azimuthal backflow in light carrying orbital angular momentum. *Optica* **10,** 1217 - 1222 (2023).

[27] Murphy, L. R.,&Bird, D. Azimuthal confinement: the missing ingredient in understanding confinement loss in antiresonant, hollow-core fibers. *Optica* **10**, 854 - 870 (2023).

[28] Li, W., Yang, S., Chen, Y., Li, C., Wang, Z. Tesla valves and capillary structures-activated thermal regulator. *Nat. Commun*. **14**, 3996 (2023).

[29] Huang, X., Anufriev, R., Jalabert, L., Watanabe, K., Taniguchi, T. et. al. A graphite thermal Tesla valve driven by hydrodynamic phonon transport. *Nature* **634**, 1086 - 1090 (2024).

[30] Pryamikov, A. D. Local Energy Velocity of the Air-Core Modes in Hollow-Core Fibers. *Photonics* **10**, 1035 (2023).

[31] Petrovich, M., Fokoua, E. N., Chen, Y., Sakr, H., Adamu, et. al. Broadband optical fibre with an attenuation lower than 0.1 decibel per kilometer. *Nat. Photon.* (2025). https://doi.org/10.1038/s41566-025-01747-5

# Supplementary

In order to correctly determine the loss mechanisms for the air - core modes of hollow – core fibers with a "negative" core boundary curvature, the following simple mathematical model for the leakage of air - core mode energy can be proposed for consideration. It is obvious that in the case of a stationary problem the following equation for the Poynting vector of the leaky air - core mode $\vec{P}(r,\varphi,z)$ is true:

$$div\vec{P} = 0 .$$

Taking into account the dependence of each of the components of the Poynting vector of the air - core mode on the z coordinate in the form of $e^{-2\operatorname{Im}(\beta)z}$, where $\operatorname{Im}(\beta)$ is the imaginary part of the propagation constant of the air - core mode ($\beta = 2\pi n_{eff} / \lambda$, where $n_{\text{eff}}$ is an effective refractive index of the air – core mode and λ is a wavelength), it is possible to obtain an expression determining the leakage losses for a given mode in cylindrical coordinates:

$$2\operatorname{Im}(\beta) P_z = div\vec{P}_{transv}, \quad (1)$$

where $\vec{P}_{transv} = \{P_r; P_\varphi\}$. We take the integral of both parts of equality (1) over a small volume $V$, assuming that the imaginary part $\operatorname{Im}(\beta)$ is small and the component $P_z$ can be considered as a weakly changing function in this volume. In this case, according to the Gauss theorem, one obtains the following expression for the imaginary part of the mode propagation constant:

$$\operatorname{Im}(\beta) = \oiint \left(P_r\vec{r} + P_\varphi\vec{\varphi}\right)\vec{n}dS / \left(2VP_z\right), \quad (2)$$

where $dS$ is the area element of the curved surface of the air – core of the hollow – core fiber and $\vec{n}$ is the normal to this surface. The value of leakage losses is determined by the value $\operatorname{Im}(\beta)$ and is constant for constant dimensions of the cross-section of the hollow – core fiber. In a hollow - core circular fiber, such as a capillary, the energy flux of the core boundary is determined only by the term $P_r$ in expression (2) (Fig. 1(left)). When introducing "negative curvature" for the core boundary of a leaky hollow – core fiber (Fig. 1(right)), a $P_\varphi$ component should arise in those regions of the core boundary where $(\vec{r}\vec{n}) \to 0$ and $(\vec{\varphi}\vec{n}) \to 1$ due to the continuity of the energy flow $\Phi = (P_r\vec{r} + P_\varphi\vec{\varphi})\vec{n}dS$. Thus, from the point of view of continuity of the outgoing energy flow of the air - core mode, the streamline of its Poynting vector should turn along the curved core boundary (Fig. 2). On the lateral surfaces of the cladding elements, the values of the radial component of the Poynting vector should decrease and the value of $P_\varphi$ should increase. In the areas of the cross-section of the hollow – core fiber between the walls of the semi-capillaries (Fig. 1(right)), two oppositely directed azimuthal flows of the air - core mode energy arise in directions opposite to each other, which leads to the formation of vortex movements of the Poynting vector. The more the curvature of the surface of the hollow fiber cladding element changes, reducing the radial component of the air - core mode energy flux, the more vortex motions arise between the walls and in the walls of these elements. Thus, it is obvious that in the presence of vortices of energy flow the components of $P_r$ and $P_\varphi$ can be alternating functions along the air - core boundary, which, when integrating expression (2), can significantly reduce the magnitude of leakage losses in contrast to a hollow - core fiber with similar geometric parameters but with a round core boundary.

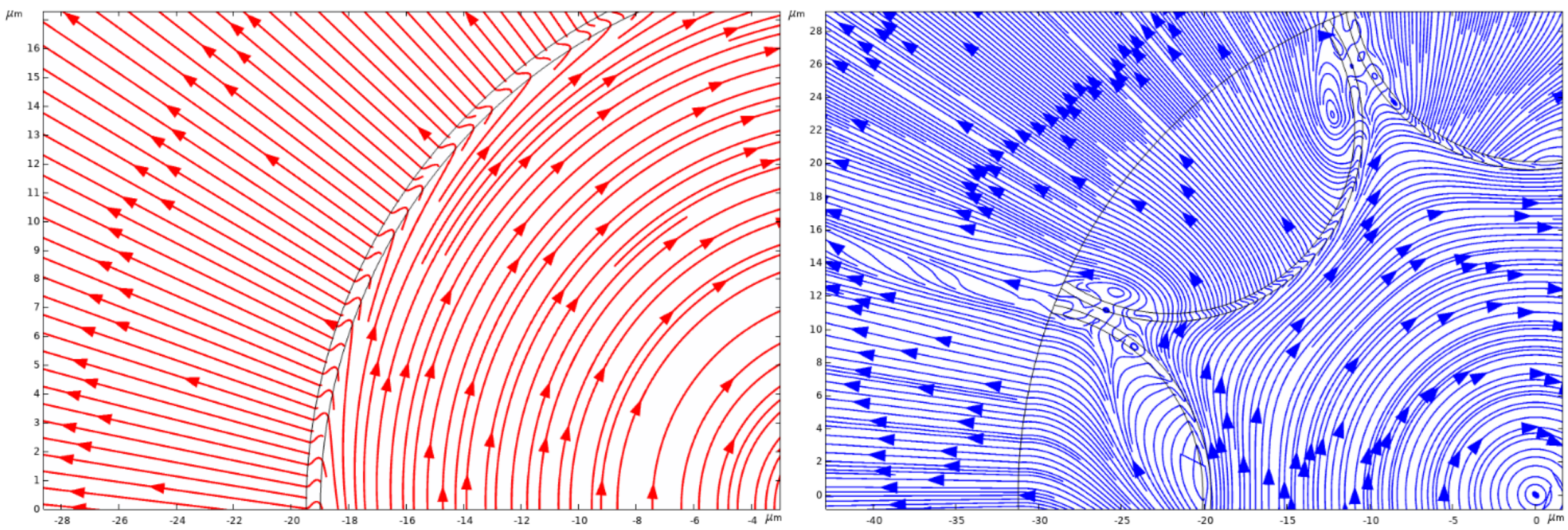

Figure 1: (left) The streamlines lines of the Poynting vector for the fundamental air - core mode of a hollow – core fiber in the form of a silica glass capillary with a core diameter of $D_{core}$ = 38 µm and a wall thickness of 490 nm at wavelength of 1.55 µm; (right) the streamlines lines of the Poynting vector for the fundamental air - core mode of a hollow – core fiber with a cladding consisting of 8 semi-capillaries, the geometric parameters are the same as for a single silica glass capillary.

The vortex motion of the air - core mode energy in the walls of the core boundary generates reverse energy flows directed along the normal to the core boundary with "negative" curvature. Thus, the energy flows directed along the normal are positive and the outgoing mode energy flow is negative. Since the amount of energy flowing out is always greater than the amount of energy flowing back into the air - core in formula (2), it is always necessary to put a minus sign in front of the expression on the right for hollow light guides with negative curvature of the core boundary.

The concept of the emergence and behavior of reverse energy flows of air - core modes at the core boundary can be clearly demonstrated using the example of energy flows of the fundamental and second modes in the horizontal direction for a hollow fiber with a negative curvature of the core - cladding boundary shown in Fig. 1(right). The leakage losses for the fundamental and second modes differ by a factor of 8 at a wavelength of 1.55 μm.

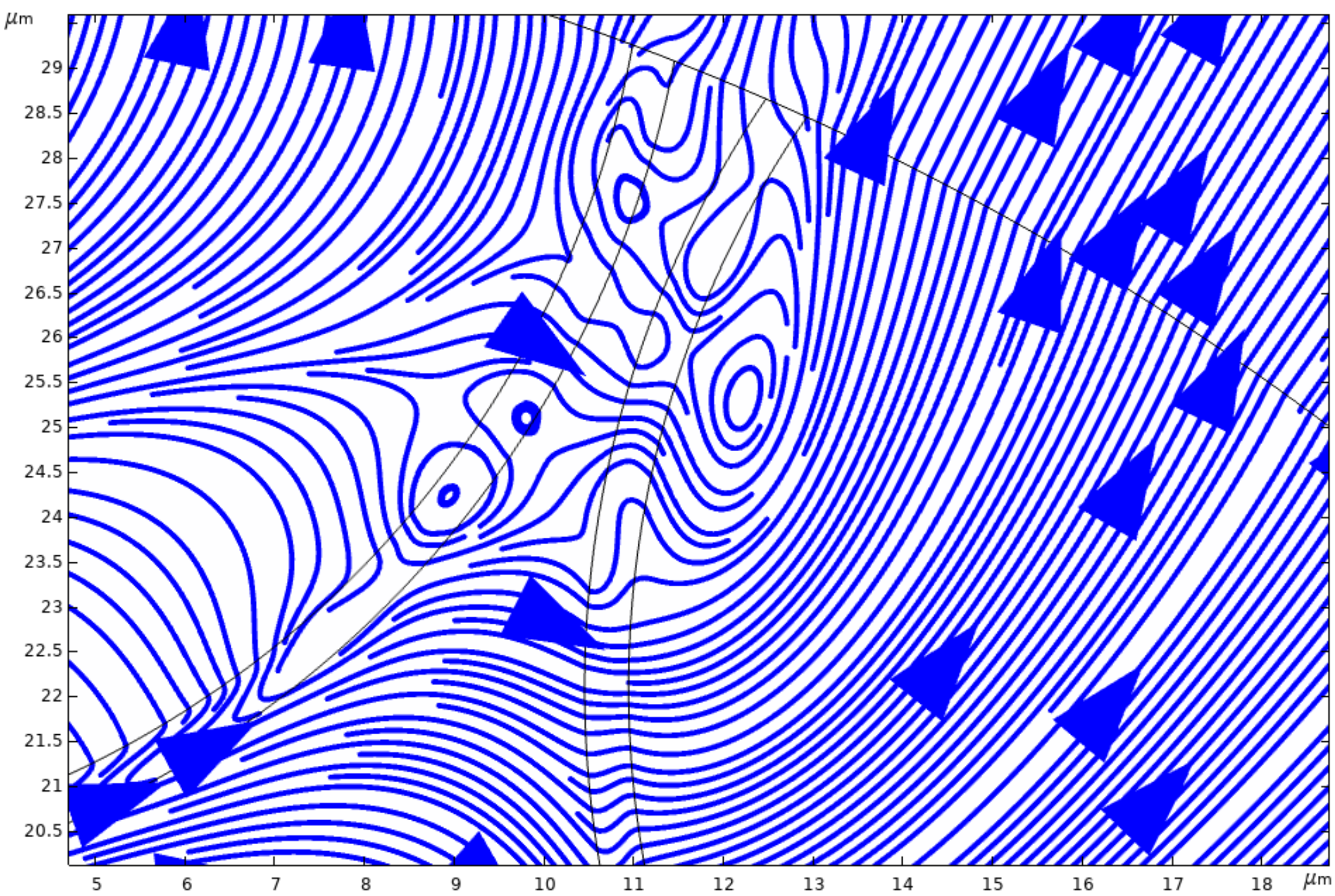

Fig. 2. Streamlines of the transverse component of the Poynting vector for the fundamental air core mode for a hollow – core fiber Fig. 1 (right), enlarged image of the cross-sectional region between the two cladding half-capillaries.

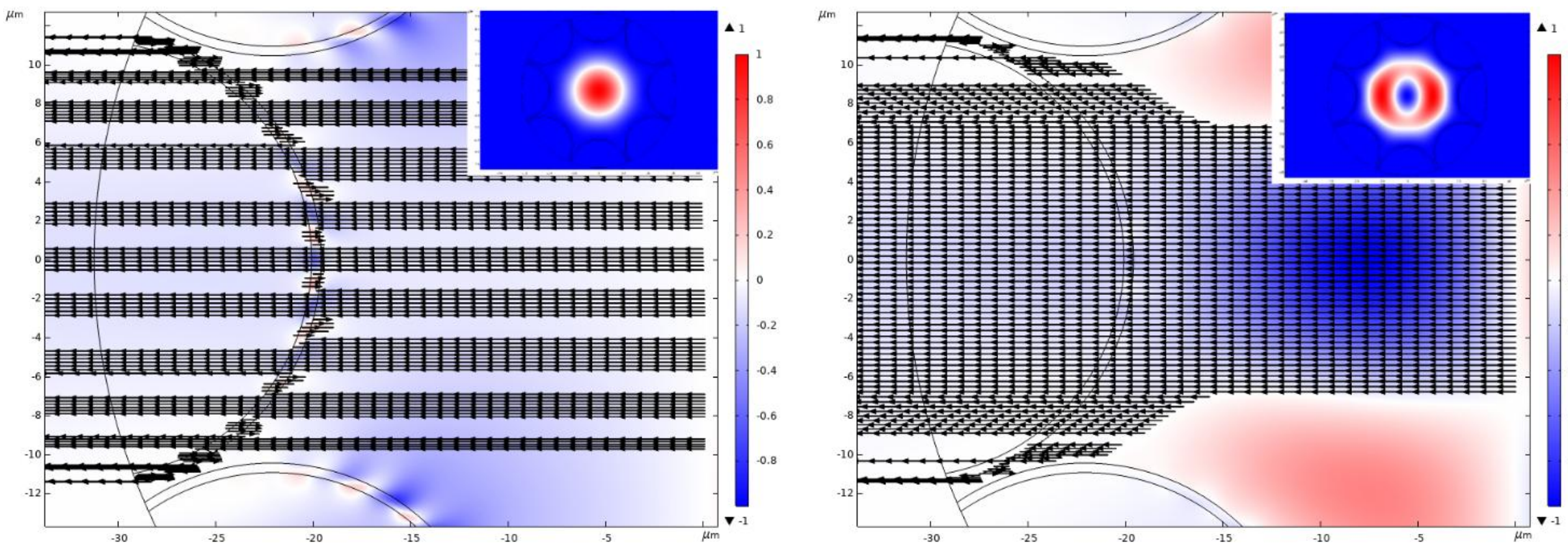

Fig. 3. Energy flux of air - core modes in the negative direction of x-axis for a hollow – core fiber shown in Fig. 1 (left) for fundamental air – core mode; (right) for the second air - core mode. The color shows the distribution of $P_x$ component of the Poynting vector over the cross section of the fiber.

From Fig. 3 it is evident that in the case of the fundamental air - core mode, clearly expressed reverse energy flows arise in the wall of the cladding semi-capillary, which have a positive sign, as indicated above. In the case of the second air - core mode, reverse energy flows occur only at the base of the cladding semi-capillary. If we take the average value of the integral for $P_x$ component of the Poynting vector along the boundary of the semi capillary in both cases, they will differ by approximately 7 times. Thus, the presence of reverse energy flows for the air - core modes significantly affects the process of mode energy outflow from the air core.

Hollow – core fibers are leaky so the electric and magnetic fields of the hollow core modes are complex analytic fields. This means that the curves describing the zeros of the imaginary and real parts of these functions can either overlap or intersect in certain areas of the cross section of the fiber. This can lead to the appearance of phase dislocations of the air - core mode fields which may also be associated with the emergence of singularities of the Poynting vector of this mode.

The presence of pronounced phase dislocations for the components of the vector fields of the air – core modes is also associated with the presence of "negative" curvature of the core boundary. Let us consider the appearance of phase dislocations using the specific example of two hollow – core fibers, considered earlier (Fig. 1). The operating wavelength is 1.55 μm. The phase distributions for the azimuthal component of the electric field $E_\varphi$ of the fundamental air - core mode for both hollow - core fibers are shown in Fig. 4.

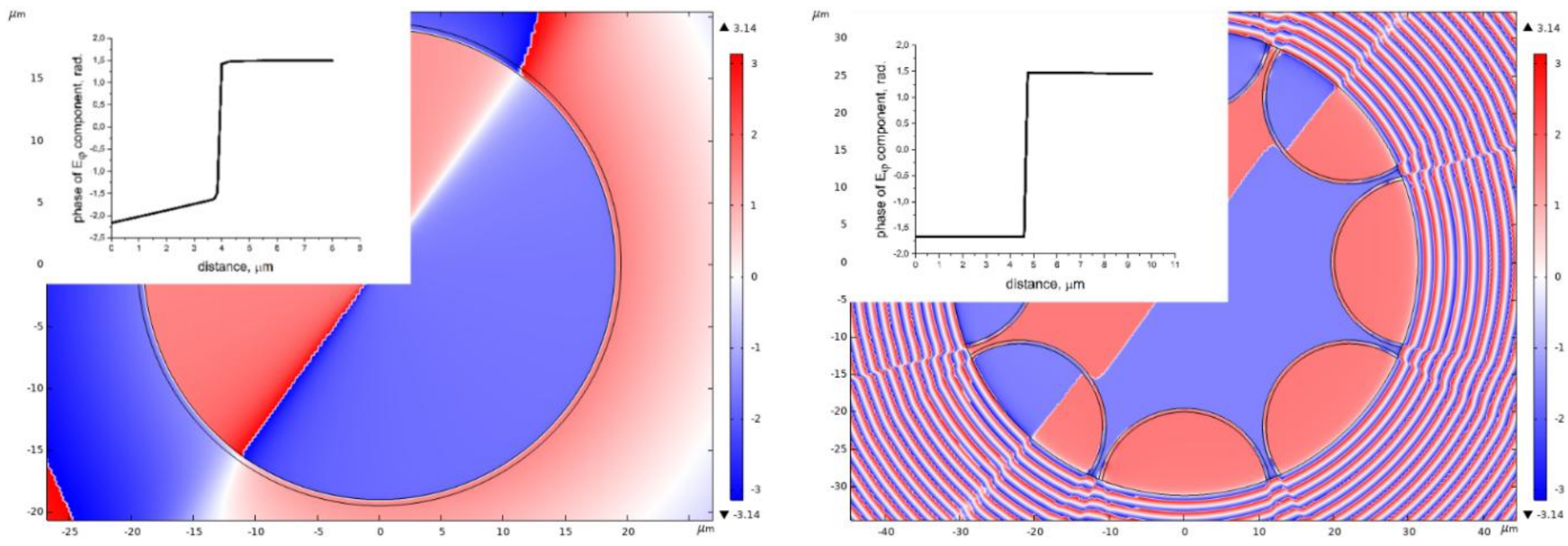

Figure 4: (left) The color distribution is the phase for the component $E_\varphi$ of the electric field of the fundamental air - core mode for silica glass capillary with $D_{core} = 38$ μm and wall thickness of 490 nm at wavelength of 1.55 μm, and the inset shows a phase jump that occurs when passing through the capillary wall; (right) the same distribution for the hollow – core fiber with negative curvature core – cladding boundary with the same geometric parameters, the inset shows a phase jump equal to π that occurs when passing through the wall of the cladding semi capillary.

When a phase dislocation (phase edge dislocation) is formed, the phase jump must be exactly equal to $\pi$ radians. In this case, the field components of the air - core vector mode will oscillate exactly in antiphase on both sides of the cladding capillary wall. Since the wavelength of 1.55 μm is related to the longest wavelength transmission band of hollow core fibers (Fig. 1), the phase jump equal to π when passing through the core - cladding boundary is one. In this case, there is only one phase dislocation $\mathrm{Re}\left(E_\varphi\right) = \mathrm{Im}\left(E_\varphi\right) = 0$ in the wall of the capillary or semi-capillary. This is exactly the case observed in the case of a hollow - core fiber with a curved core boundary (Fig. 4 (right)). In the case of a single - capillary with a round core boundary (Fig. 4 (left), the azimuthal component of the electric field of fundamental air core mode is quite strongly out of phase on both sides of the capillary wall. This means that for this shape of the core boundary of the hollow – core fiber, there is no exact phase dislocation. Usually, it is the formation of phase dislocations of one type or another for the components of the vector mode fields that leads to the formation of singularities of the Poynting vector and the corresponding vortex motions of the radiation energy of these modes [1].

Based on the above, to reduce losses in hollow – core fibers, it is necessary to optimize the configuration of the Poynting vector vortices of the air - core mode and, accordingly, its reverse energy flows in the cladding capillaries. To do this, it is possible to change the shape of the surface of the cladding capillaries without inserting too many additional capillaries into them [2]. For example, a hollow – core fiber with a cladding consisting of six elliptical capillaries with six more elliptical capillaries inserted into them (Fig. 5).

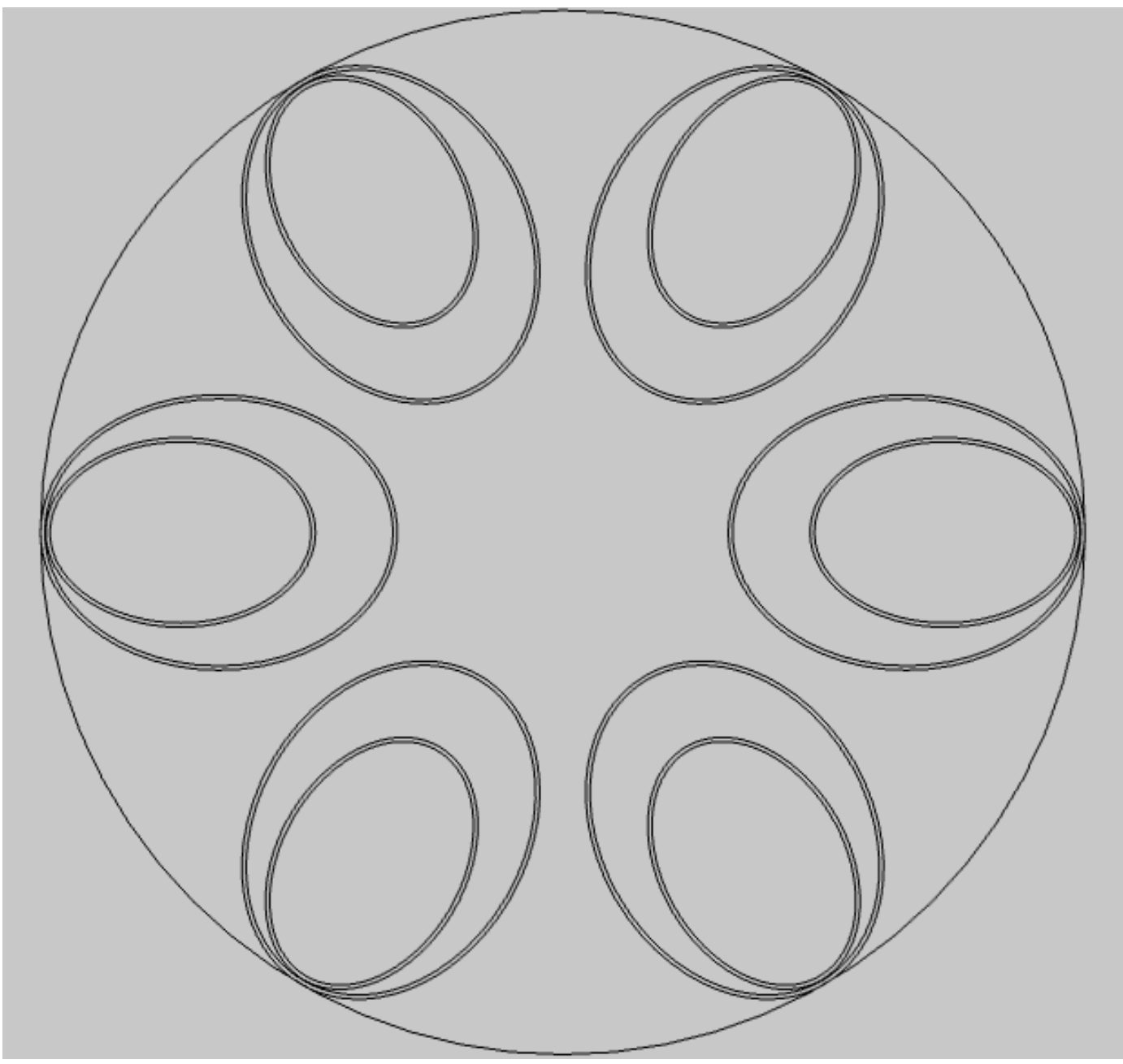

Fig. 5. Nested silica glass hollow – core fiber with 12 elliptical capillaries in the cladding. The air – core diameter is 30 μm, the cladding capillary wall thickness of 380 nm. The ratio of the minor semi-axis of the ellipse to the major one for a large capillary is 0.77 and for a small one is 0.7.

The losses of the fundamental air - core mode in such a hollow – core fiber are 0.02 dB/km at a wavelength of 1.55 μm, and for the second air - core mode they are 22 dB/km. The leaky energy flows of the air - core modes in the negative direction of the x-axis are shown in Fig. 6. From Fig. 6 it is evident that reverse energy flows are observed for both air - core modes since the mode losses are small, but in the case of the fundamental mode their configuration is much denser on the wall of the inserted cladding capillary.

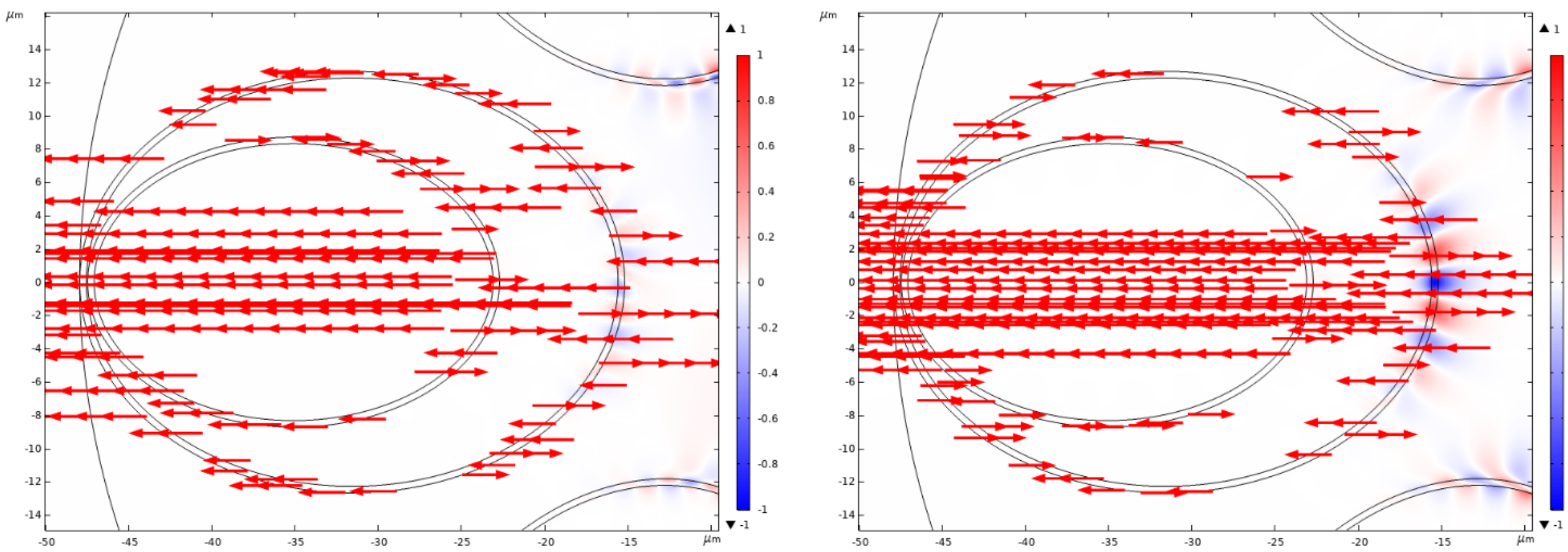

Fig. 6. . Energy flux of air - core modes in the negative direction of x-axis for a hollow – core fiber shown in Fig. 5 (left) for fundamental air – core mode; (right) for the second air - core mode. The color shows the distribution of $P_x$ component of the Poynting vector over the cross section of the fiber.

Thus, by varying only the shape of the surface of the cladding capillaries, it is possible to change the configuration of the reverse energy flows of the air - core modes, thereby changing the level of their losses.

## References

[1] Novitsky, A. V. & Barkovsky, L. M. Poynting singularities in optical dynamic systems. *Phys. Rev. A* **79**, 033821 (2009).

[2] Petrovich, M., Fokoua, E. N., Chen, Y., Sakr, H., Adamu, et. al. Broadband optical fibre with an attenuation lower than 0.1 decibel per kilometer. *Nat. Photon.* (2025). https://doi.org/10.1038/s41566-025-01747-5